\documentclass[preprint]{aastex631}

\newcommand{\teff}{$T_{\rm eff}$}
\newcommand{\rstar}{$R_\star$}
\newcommand{\chz}{CHZ$_2$}

\shorttitle{}
\shortauthors{Ware et al.}

\begin{document}

\title{Continuous Habitable Zone Metric for Prioritizing Habitable Worlds Observatory Targets}

\author[0000-0003-2828-0334]{Austin Ware}
\affiliation{School of Earth and Space Exploration \\
Arizona State University \\
Tempe, AZ 85287, USA}

\author[0000-0003-1705-5991]{Patrick Young}
\affiliation{School of Earth and Space Exploration \\
Arizona State University \\
Tempe, AZ 85287, USA}



\begin{abstract}

Future direct imaging space telescopes, such as NASA's Habitable Worlds Observatory (HWO), will be the first capable of both detecting and characterizing terrestrial exoplanets in the habitable zones (HZ) of nearby Sun-like stars. Since this will require a significant amount of time and resources for even a single system or exoplanet, the likelihood that a system will host detectable life should be considered when prioritizing observations. One method of prioritization is to estimate the likelihood that an exoplanet has remained continuously within the HZ long enough for life to emerge and make a detectable impact on the atmosphere. We utilize a Bayesian method to calculate the likelihood that a given orbital radius around a star is currently in the 2 Gyr continuous habitable zone (\chz), the approximate time it took life on Earth to significantly oxygenate the atmosphere. We apply this method to the 164 stars in the NASA Exoplanet Exploration Program Mission Star List (EMSL) for HWO, representing a preliminary sample of Sun-like stars with HZs most accessible to a future direct imaging mission. By considering the \chz\ likelihood at all orbital radii outside a hypothetical inner working angle for HWO, we define a metric for prioritizing targets according to the accessibility and total extent of the \chz. We find that the \chz\ metric peaks between $3-4$ Gyr for late-F and early-G dwarfs, but tentatively determine that stars earlier than $\sim {F3}$ or hotter than $\sim 6600$ K are unlikely to have a \chz\ at the time of observation.

\end{abstract}



\section{Introduction} \label{ch4:sec:intro}

At present, opportunities are limited for detecting, characterizing, and searching for the signatures of life in the atmospheres of terrestrial habitable zone (HZ) exoplanets. The \textit{James Webb Space Telescope} (\textit{JWST}) is restricted to transiting exoplanets in the HZs of the nearest and lowest-mass M dwarfs. To access the atmospheres of terrestrial exoplanets in the HZs of Sun-like (F-, G-, K-type) stars with any orbital inclination, direct imaging of reflected light is required \citep{2019arXiv191206219T,2020arXiv200106683G}. However, for an Earth-twin orbiting in the HZ of a G2-dwarf, detection of reflected light requires a sensitivity of $\sim 10^{-10}$ for the planet-star contrast at visible wavelengths \citep{2018AsBio..18..739F}, well beyond the capabilities of current instruments. The future NASA's Habitable Worlds Observatory (HWO) will aim to be the first telescope capable of achieving this feat and fulfill the Astro2020 Decadal Survey recommendation of characterizing and searching for biosignatures in the atmospheres of $\sim 25$ terrestrial HZ exoplanets. Reaching this level of contrast and performing detailed spectroscopic characterization of detected terrestrial HZ exoplanets will require a significant amount of time and resources for even a single system. This makes prioritization based on the likelihood that discovered exoplanets will host detectable life critical to consider.

With the release of the NASA ExEP mission star list (EMSL) for HWO \citep{2024arXiv240212414M}, we now have a preliminary sample of stars with HZs most accessible for a future direct imaging survey with a 6-m-class telescope. Taking into account the conservative moist and maximum greenhouse HZ limits for Earth \citep[0.95-1.67 AU,][]{1993Icar..101..108K,2013ApJ...765..131K} scaled by the luminosity for other stars, the estimated exoplanet brightness and planet-star contrast ratio, the presence of dust disks and/or close-in binaries, and hypothetical IWAs for HWO, the EMSL reports a total of 164 potential targets. These targets are assigned to Tiers A, B, and C based on the constraints used, with Tier A stars meeting the strictest constraints. However, the effects of stellar effective temperature (\teff) on the HZ, alternative HZ models, or the evolution of the HZ, which allows for further prioritization based on the potential for habitability, are not considered. \teff\ will have a significant effect on the extent of the HZ \citep[e.g.,][]{2013AsBio..13..251R,2013AsBio..13..715S,2015ApJ...806..137R} and 1D and 3D climate models have shown a wide range of possible HZ limits \citep[e.g., see][and references therein]{2017ApJ...837..107W}.

As of the beginning of 2025, the HWO Target Stars and Systems Working Group has adopted the EMSL as the highest priority "Tier 1" target list for HWO. Two additional and more expansive tiers were also adopted and contain lower priority targets that may still be considered for observations with HWO. "Tier 2" contains $\sim 500$ targets selected through Exo-Earth yield simulations with the Altruistic Yield Optimization \citep[AYO,][]{ayo} and Exoplanet Open-Source Imaging Mission Simulator \citep[EXOSIMS,][]{exosims} tools. "Tier 3" contains $\sim 13,000$ nearby ($<50$ pc), bright ($T<12$ mag) stars and was adopted from the HWO Preliminary Input Catalog \citep[HPIC,][]{hpic}. While we focus on the more manageable set of 164 stars in the EMSL here, future work will include a larger input catalog that will be more representative of all possible targets HWO may observe.

The EMSL is composed mainly of F-, G-, and K-type stars, where HZ evolution with respect to stellar luminosity plays a more significant role over billion-year time frames. This presents an opportunity to demonstrate how the Bayesian continuous habitable zone (CHZ) method from \citet{2022ApJ...929..143W} can be used to further prioritize targets based on the potential of continuous habitability. The CHZ has been formulated in several ways in the literature, but is typically defined as either the sustained CHZ or a fixed-duration CHZ. The sustained CHZ is defined as the orbital area that has remained in the HZ from the zero-age main sequence until the current age of the star \citep[e.g.,][]{tuchow2020}. The fixed-duration CHZ neglects whether a planet originated in the HZ and is instead defined as the orbital area that remains the HZ for a set amount of time \citep[e.g.,][]{2020AJ....159...55T}. In this work, we implement the fixed-duration 2 Gyr CHZ (\chz) definition from \citet{2020AJ....159...55T}. This HZ lifetime corresponds to when the Great Oxidation Event (GOE) occurred on Earth, about 2 Gyr after formation, and denotes the approximate time it took life on Earth to have a significant detectable impact on the atmosphere \citep[e.g.,][]{doi:10.1098/rstb.2006.1838}.

With a grid of stellar evolution models, we can evolve the HZ over time for each model and calculate the inner and outer edges of the \chz\ for each time step. Using the measured stellar properties as inputs, we can then implement a sampling algorithm, such as the Markov Chain Monte Carlo (MCMC) method, to sample the joint posterior distribution for the locations of the inner and outer edges of the \chz. The joint posterior sampling can then be converted to a \chz\ posterior likelihood distribution by calculating the frequency with which each orbital radius is within the limits of the \chz. The total extent of the \chz\ observable by HWO can then be used as a factor in prioritizing targets according to the likelihood that planets host detectable life.

We apply this method to the 164 stars in the EMSL, combining an updated grid of Tycho stellar evolution tracks with the runaway greenhouse and maximum greenhouse HZ prescriptions of \citet{2014ApJ...787L..29K}. In addition, we incorporate the inner HZ (IHZ) prescription from \citet{2023AA...679A.126T}, which defines the initial IHZ boundary where Earth-like planets will be able to form oceans following formation. As mentioned, we consider here the total extent of the \chz\ observable by HWO. For targets in the EMSL, which lack discovered terrestrial HZ exoplanets, we cannot calculate the \chz\ likelihood for individual exoplanets as done in \citet{2022ApJ...929..143W}. Instead, we derive a \chz\ metric by integrating the \chz\ posterior likelihood distribution, with respect to orbital radius, outside a hypothetical inner working angle for the HWO coronagraph \citep{2024arXiv240212414M}. The \chz\ metric then enables direct comparisons between stars according to the \chz\ accessibility. \citet{tuchow2020} previously explored a similar metric for prioritizing targets for future missions according to the continuous habitability potential of exoplanets. The authors investigated how various formulations for the CHZ and models of planet occurrence rate and biosignature production would impact the estimated biosignature yields for targets. The metric proposed in this work differs mainly in that we do not consider the planet occurrence rate or biosignature production, instead focusing on the \chz\ accessibility for an HWO-like mission within a Bayesian framework.

To streamline the process of determining stellar ages and \chz\ posterior likelihood distributions compared to \citet{2022ApJ...929..143W}, we combine both calculations inside the open-source stellar model grid interpolation code \texttt{kiauhoku} \citep{2020ApJ...888...43C}. This enables us to determine stellar ages, masses, and \chz\ posteriors simultaneously, greatly improving the efficiency and reproducibility of our results. As inputs for our calculations, we adopt \teff\ and stellar radii (\rstar) from the SPORES (Stellar/System Properties \& Observational Reconnaissance for Exoplanet Studies with HWO) catalog \citep{2024ApJS..272...30H}. The SPORES catalog reports uniformly derived \teff\ and \rstar\ from fits to spectral energy distributions (SED) for the EMSL sample. The catalog also contains masses and ages from fits to stellar isochrones, which we use as a comparison for our results.

In Section \ref{ch4:sec:meth}, we discuss updates to the Tycho stellar evolution code and model grid, the HZ prescriptions used, our updated Bayesian \chz\ method, and the process for calculating the \chz\ metric. Section \ref{ch4:sec:res} reports our results and compares our masses and ages with those in the SPORES catalog. Section \ref{ch4:sec:disc} discusses results and trends for the \chz\ metric and future work. Finally, Section \ref{ch4:sec:conc} presents a summary of our work.

\section{Methods}\label{ch4:sec:meth}

\subsection{Tycho Stellar Evolution Catalog}\label{ch4:sec:tycho}

We use the stellar evolution code Tycho \citep{2005ApJ...618..908Y} to generate a new catalog of stellar evolution tracks. Tycho is a 1D stellar evolution code that utilizes a hydrodynamic formulation of the stellar evolution equations. We have updated Tycho to include the latest Los Alamos OPLIB high-temperature opacities \citep{2016ApJ...817..116C} and \AE SOPUS 2.0 low-temperature opacities \citep{2022ApJ...940..129M}, a 522 element reaction network up to $^{99}$Tc utilizing the latest REACLIB rates \citep{2010ApJS..189..240C}, and the \citet{2021AA...653A.141A} protosolar abundances as our reference solar mixture. As with previous versions, Tycho uses a combined OPAL and Timmes equation of state \citep{1999ApJS..125..277T,2002ApJ...576.1064R}, gravitationally induced diffusion \citep{1994ApJ...421..828T}, general relativistic gravity, time lapse, curvature, and automatic rezoning. The reaction network uses weak rates from \citet{2000NuPhA.673..481L} and screening from \citet{1973ApJ...181..457G}. Neutrino cooling and mass loss are included, but the latter is trivial for the low-mass stars considered in this work. Tycho avoids the need for free convective parameters (i.e., "convective overshoot") by using a description of turbulent convection based on 3D, well-resolved simulations of convection between stable layers \citep{2007ApJ...667..448M,2009ApJ...690.1715A,2010ApJ...710.1619A,2011ApJ...733...78A}.

The new Tycho stellar evolution catalog spans masses of $0.2-1.6$ M$_\odot$ in steps of 0.01 M$_\odot$, metallicities ([M/H]) of $-1.75-0.5$ dex in steps of 0.25 dex, and evolutionary phases from the pre-main sequence birth to the first convective dredge-up. These parameter ranges are sufficient for studying the 164 F, G, K and early-M main sequence and subgiant stars in the EMSL, with the caveat that we neglect rotation and departures from the initial scaled solar abundances.

\subsection{SPORES Catalog}\label{ch4:sec:spore}

Following the publication of the EMSL for HWO, the SPORES collaboration created a catalog of system properties for all 164 stars \citep{2024ApJS..272...30H}. This includes averaged spectroscopic abundances from the Hypatia Catalog \citep{2014AJ....148...54H} and uniformly derived stellar properties via SED modeling. We adopt the reported \teff, \rstar, and iron abundances ([Fe/H]) for use in Section \ref{ch4:sec:kiauhoku} to calculate stellar ages, masses, and \chz\ posterior likelihood distributions. The [Fe/H] values reported in the SPORES Catalog are normalized to the solar photospheric iron abundance reported in \citet{2009ARAA..47..481A}. We normalize the photospheric [Fe/H] for each timestep in the Tycho stellar evolution tracks to the same scale as the SPORES catalog to avoid introducing systematic errors when fitting to the Tycho models. For stars in the SPORES catalog without [Fe/H] from Hypatia, we use the values reported in the EMSL. The SPORES catalog also reports masses and ages from sampling of the MIST (MESA Isochrones and Stellar Tracks) isochrone grid \citep{2016ApJS..222....8D}, which we compare to in Section \ref{ch4:sec:ages_masses}. We note that the calculation of masses and ages was not a primary focus in the construction of the SPORES catalog, but rather a bonus of the SED fitting package used: \textit{ARIADNE} \citep[spectrAl eneRgy dIstribution bAyesian moDel averagiNg fittEr,][]{ariadne}. Therefore, the masses and ages reported in the SPORES catalog should be seen as relatively low fidelity estimates. As the goal of this work is to demonstrate the utility of the \chz\ metric, we use the SPORES catalog as a simple consistency check with a uniformly derived set of values specific to the EMSL.

\subsection{Habitable Zone Models}\label{ch4:sec:hz_mod}

Following \citet{2022ApJ...929..143W}, we include the 1 M$_\oplus$ HZ model from \citet{2014ApJ...787L..29K}, defined by a runaway greenhouse IHZ and maximum greenhouse OHZ. The authors parameterize the HZ with respect to the $T_{\rm eff}$ and luminosity of the star. By taking outputs from the Tycho stellar evolution models, we calculate the HZ boundaries for each time-step of the star's main sequence evolution.

\begin{figure}[htb!]
    \centering
    \includegraphics[width=\textwidth]{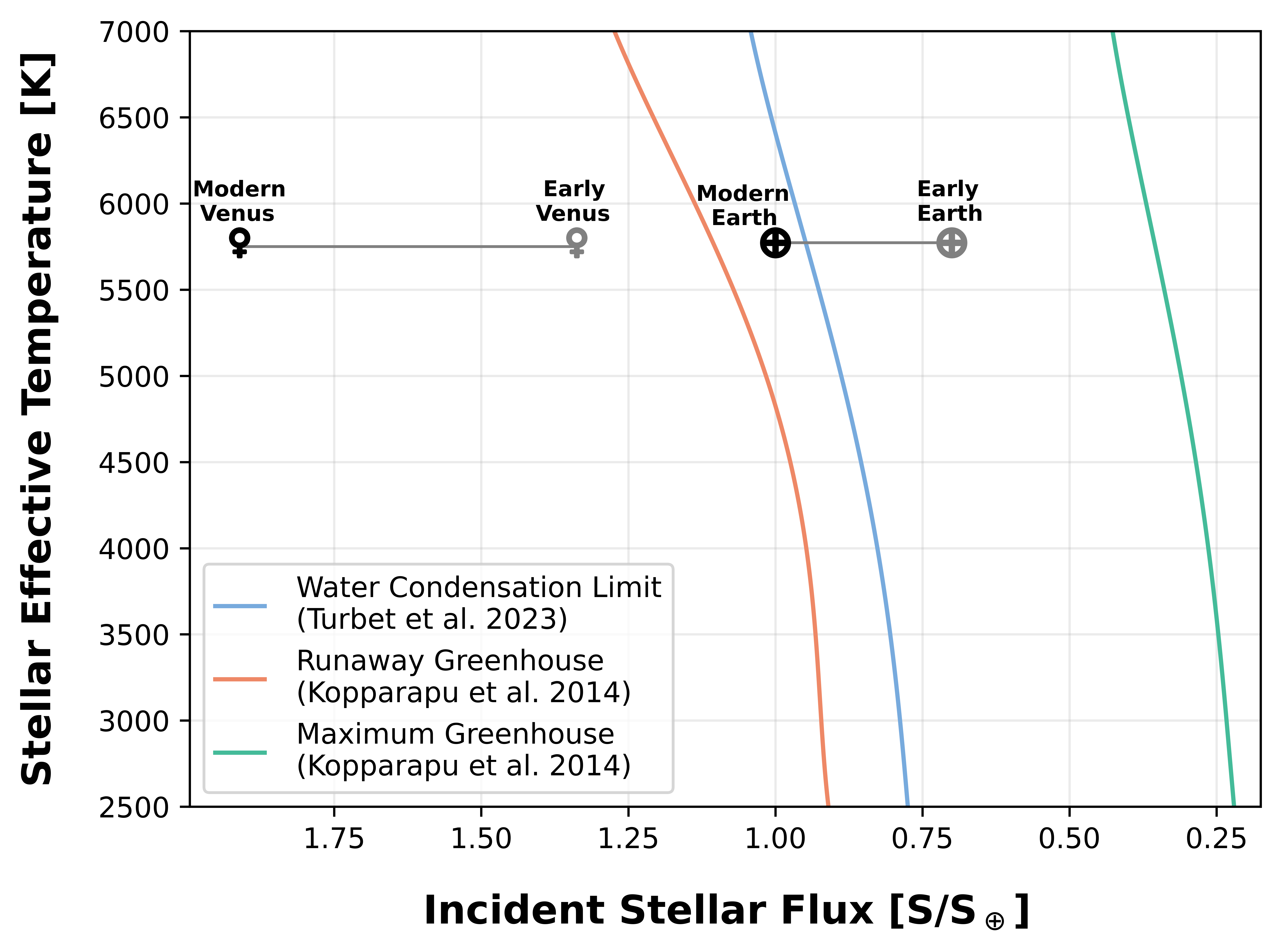}
    \caption{HZ prescriptions used in this work: runaway greenhouse IHZ and maximum greenhouse OHZ from \citet{2014ApJ...787L..29K}; water condensation limit IHZ from \citet{2023AA...679A.126T}. Over-plotted are the instellations for Venus and Earth at the ZAMS and solar age (4.57 Gyr).}
    \label{ch4:fig:hz_mods}
\end{figure}

We also include a new IHZ prescription from \citet{2023AA...679A.126T}, determined using a 3D general circulation model (GCM), as a compliment to the 1D HZ model from \citet{2014ApJ...787L..29K}. Climate models often assume the planet’s initial water begins condensed on the surface. \citet{2021Natur.598..276T} showed that if the water is initially assumed to be fully vaporized following the magma ocean phase, a new IHZ instellation limit can be determined where atmospheric water vapor condenses on the surface. The authors find that this so-called "water condensation limit" (WCL) is always at a lower instellation than the runaway greenhouse limit at the zero-age main sequence (ZAMS). If a planet has an instellation exceeding the WCL at the ZAMS, that planet will be unable to ever form oceans on its surface and will lose its water through photolysis and atmospheric escape. This has significant consequences for planets near the IHZ, such as Venus. Figure \ref{ch4:fig:hz_mods} shows the HZ prescriptions along with the instellations for Venus and Earth at the ZAMS and at solar age (4.57 Gyr). While Earth would have been outside the WCL at the ZAMS, Venus has always been inside the WCL. This implies Venus would never have been able to form oceans. We incorporate these results into our model, with the WCL defining the initial IHZ at the ZAMS. We keep the IHZ distance constant as the star evolves until the stellar luminosity increase causes the runaway greenhouse limit to exceed the WCL.

\subsection{Stellar Masses, Ages, and \chz\ Posterior Likelihood Distributions}\label{ch4:sec:kiauhoku}

In \citet{2022ApJ...929..143W}, we calculated stellar ages and \chz\ posterior likelihood distributions separately. This involved first performing $\chi^2$ fits to Tycho stellar evolution tracks to obtain best-fit stellar ages and/or retrieving literature measurements of stellar age determined via empirical age relations. We then used the derived age, combined with measured stellar masses and metallicities from the literature, to place prior probabilities on our HZ evolution models. By calculating the Bayesian likelihood that a range of orbital radii are within the \chz\, we produced posterior likelihood distributions for the location of the \chz\ around a given star. While this method is functional and effective, we have significantly improved the efficiency and reproducibility of our method by combining both calculations within the open-source \texttt{kiauhoku}\footnote{\url{https://github.com/zclaytor/kiauhoku}} stellar model grid interpolator.

First presented in \citet{2020ApJ...888...43C}, \texttt{kiauhoku} performs interpolation between stellar evolution tracks that have been transformed to an equivalent evolutionary phase (EEP) basis (see Appendix \ref{sec:eep} for additional information) and maps stellar observables, in this case \teff, \rstar, and [Fe/H], onto the stellar model fundamental properties of mass, metallicity, and age. Combined with the MCMC sampling Python package \texttt{emcee} \citep{2013PASP..125..306F,2019JOSS....4.1864F}, this results in sampled posterior distributions for each parameter of interest. In order to include the location of the \chz\ as a sampled parameter, we calculate the \chz\ boundaries for each timestep in the stellar evolution tracks, starting from the ZAMS. We then use an ensemble of walkers running individual Markov chains to sample the posterior distributions for mass, age and \chz\ location.

We place priors on the initial mass, initial metallicity, age, and EEP range. For mass and metallicity, we restrict the sampling to within the bounds of the Tycho model grid and use the joint log-normal and power-law Chabrier initial mass function \citep{2003PASP..115..763C} to positively weight lower mass models. For age, we use the Gaussian Q-function to constrain samples below the age of the universe \citep[13.787$\pm$0.020 Gyr,][]{2020AA...641A...6P}, which takes the form

\begin{equation}
    P = \frac{1}{2} (1+erf(\frac{Age_{\rm mod}-Age_{\rm uni}}{\sigma_{\rm uni}\sqrt{2}}))
\end{equation}

\noindent where $erf$ is the error function, $Age_{\rm mod}$ is the model age, and $Age_{\rm uni}$ is the age of the universe. For the EEP range, we restrict the models to between the ZAMS and the beginning of the red giant branch, since all stars in the EMSL have dwarf (V) or subgiant (IV) luminosity classes. We use the luminosity classes reported in the EMSL to determine the initial guess for the EEP. After an initial MCMC run, we use the median EEP of the returned chains to re-run the MCMC sampling with the walkers distributed closer to the highest probability region. This process led us to find that the reported spectral types for several stars in the EMSL are mismatched, with luminosity classes that did not match the best-fitting evolutionary phase or the physical properties reported in both the EMSL and SPORES catalog. We discuss this further in Appendix \ref{sec:spt} and provide updated spectral types in Table \ref{ch4:tab:spt}.

We use a $\chi^2$ distribution to calculate the log-likelihood for each model grid point:

\begin{equation}
    ln(P) = -\frac{1}{2} \sum_i (\frac{x_{{\rm SPORES},i}-x_{{\rm mod},i}}{\sigma_{{\rm SPORES},i}})^2
\end{equation}

\noindent where $x_{{\rm SPORES},i}$ and $x_{{\rm mod},i}$ are the SPORES catalog and Tycho modeled \teff, \rstar, and [Fe/H], respectively. After obtaining a sampled posterior distribution for the \chz\ boundaries, we compare a range of orbital radii to the posterior distribution to determine the likelihood that any given orbital distance around a star is within the \chz.

This combined method for determining stellar ages and \chz\ locations enables us to generalize the Bayesian \chz\ routine from \citet{2022ApJ...929..143W} to any stellar model grid. This also eliminates additional uncertainty introduced by fitting ages separately and using literature masses to constrain the HZ evolution models. We have released a GitHub module within \texttt{kiauhoku} to enable users to calculate HZ and \chz\ boundaries for any stellar model grid and sample posterior distributions, including the ability to select various published HZ prescriptions, input custom HZ prescriptions via a list of polynomial coefficients, and choose any CHZ lifetime. Additionally, the Tycho stellar model grid will be included as an option to use within \texttt{kiauhoku}.

\subsection{Determining a \chz\ Metric}\label{ch4:sec:metric}

The goal of determining \chz\ posterior distributions for each star in the EMSL is to demonstrate a method for further prioritizing observations based on the likelihood a terrestrial HZ exoplanet in the system will host detectable signatures of life. However, at the time of this writing none of the EMSL star have any candidate or confirmed terrestrial HZ exoplanets, assuming an upper planet radius limit of 1.8 M$_\oplus$ and the HZ prescriptions from \citet{2014ApJ...787L..29K}. Therefore, instead of ranking exoplanets within each system based on their \chz\ likelihoods, we must derive a metric for prioritization by considering the full \chz\ posterior distribution for each system. For a star with exact knowledge of the stellar properties, such as the Sun, this metric would be defined by the total range of the \chz\, given by the inner and outer boundaries. For the stars in the EMSL with significant associated uncertainties for measured stellar properties, we must take into account the likelihood that orbital radii are within the \chz. We therefore integrate the \chz\ posterior distribution over the distance from the host star.

HWO will be a direct imaging mission, in which a coronagraph will be used to block the light from star and reveal the faint reflected light of the planets. Coronagraphs have an associated inner working angle (IWA), inside of which any planets would be hidden behind the coronagraph or the diffracted light from the host star. We must take into account the IWA when integrating over the \chz\ posterior distribution, with the IWA defining the lower integration limit. Observations will also be constrained by the outer working angle (OWA), beyond which the throughput of the coronagraph becomes insufficient. The metric equation then takes the form

\begin{equation}
    metric = \int_{IWA}^{OWA} P(CHZ_2) dr
\end{equation}

\noindent where the IWA is converted from a projected angle on the sky to an orbital radius using the distance reported in the EMSL, typically from Gaia.

We currently do not know the final specifications for HWO and must instead assume a reasonable value for the IWA. The HabEx \citep{2020arXiv200106683G} and LUVOIR \citep{2019arXiv191206219T} concept studies adopted an IWA of 2.4 $\lambda/D$ for a vector vortex charge-6 coronagraph. Assuming an observed wavelength of 1 $\mu$m, corresponding to the spectral continuum level slightly red-ward of the key 940 nm water absorption feature, and a telescope diameter of 6 meters, this predicted IWA corresponds to an on-sky angular separation of 83 mas. We adopt this value as our IWA for all stars in the EMSL, which is equivalent to the IWA adopted by the EMSL for Tier A stars. The OWA is likely to be well beyond the OHZ in all cases, so in practice we simply set the upper integration limit to infinity. However, the upper limit will be better represented by the exoplanet brightness and contrast limits of the instrument. Calculating reasonable brightness and contrast limits is far more complex, requiring additional assumptions about the telescope design, exoplanet radius, exoplanet phase angle at the time of observation, and planetary geometric albedo. This is beyond the scope of this work and we instead assume that all exoplanets outside the IWA will be observable, but we discuss the implications of this assumption further in Section \ref{ch4:sec:disc_chz}. We use these integration limits to calculate the \chz\ metric for each star. The \chz\ metric can be interpreted as a proxy for the range of the \chz\, with a higher metric indicating a combination of a more accessible \chz\ for HWO and a higher likelihood of continuous habitability for terrestrial exoplanets in the HZ.

\begin{figure}[htb!]
    \centering
    \includegraphics[width=\textwidth]{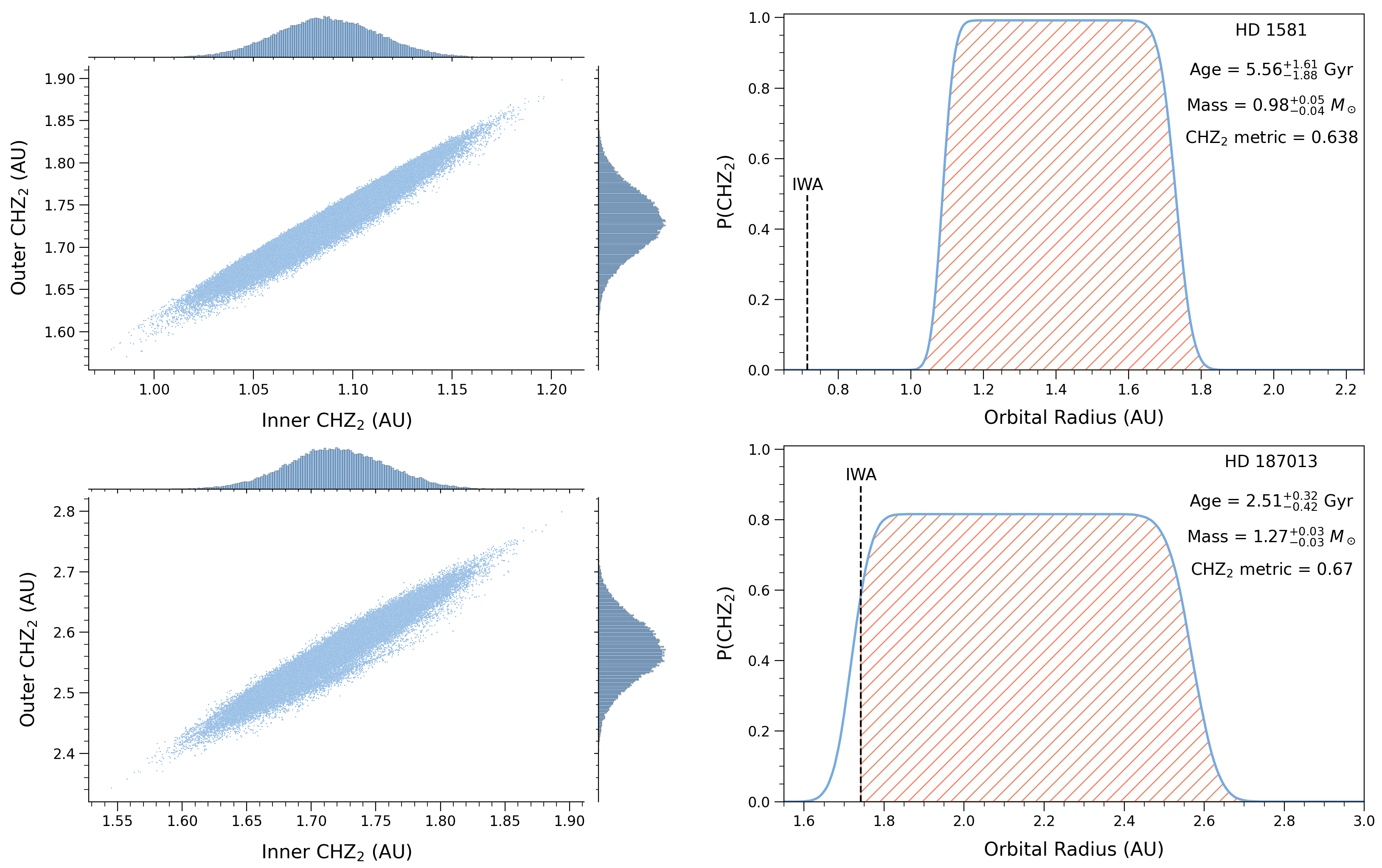}
    \caption{Left: MCMC sampled joint posterior distributions for the inner and outer \chz\ boundaries for HD 1581 (top) and HD 187013 (bottom). We note that MCMC samples for which the \chz\ does not exist are not shown, but these samples must be taken into account when calculating the \chz\ posterior likelihood distribution. Right: \chz\ posterior likelihood distributions for the same stars (blue curves). The black dashed lines denotes the orbital radius corresponding to the on-sky IWA = 83 mas. The orange hashed regions denotes the area outside the IWA that is integrated over to calculate the \chz\ metric.}
    \label{fig:jointplot}
\end{figure}

We provide visual examples of how the \chz\ posterior distribution is converted to a metric for HD 1581 and HD 187013 in Figure \ref{fig:jointplot}. First, the MCMC sampler produces a joint posterior distribution for the locations of the inner and outer \chz\ boundaries. We convert the joint posterior sampling to a \chz\ posterior likelihood distribution by calculating how often orbital radii are in the \chz. The posterior likelihood distribution is then integrated over, with respect to orbital radius, outside the IWA to calculate the \chz\ metric. We note that for HD 187013 the maximum \chz\ likelihood is $\sim 80$\%. This is due to the estimated age, $2.51^{+0.32}_{-0.42}$ Gyr, which is near the minimum age required for the \chz\ to exist. This leads to a significant number of samples where there is not yet a \chz, restricting the maximum likelihood to $< 100$\%.

\section{Results}\label{ch4:sec:res}

We used 25 walkers, a burn-in phase of 1,500 iterations, and 10,000 final iterations with the MCMC sampler for each of the 164 stars in the EMSL. We calculate the median and 68\% highest posterior density interval (HPDI) of the chains for each star. The 68\% HPDI is the shortest interval that contains 68\% of the probability mass. We report quality flags depending on how well the inputs from the SPORES catalog matched to the MCMC chains: no flag if all inputs fall within the 68\% HPDI, "1" if one or more inputs are outside the 68\% HPDI, "2" if one or more are outside the 95\% HPDI, and "3" otherwise, indicating an extremely poor sampling. We calculated the mean, median, and root median square offset (RMedS) between the sample median and the input \teff, \rstar, and [Fe/H]. These are shown in Table \ref{ch4:tab:offsets} and demonstrate that the samples are reasonably well converged, with the median and RMedS offsets being close to zero for all parameters. However, the mean offsets are slightly inflated, mostly due to 10 stars that were flagged with the lowest quality sampling. The reported \teff\ and \rstar in the SPORES catalog have median uncertainties of $0.8\%$ and $1.9\%$, respectively. These are more than two times lower than the systematic noise floors of $2.4\%$ and $4.2\%$ set by measured interferometric angular diameters and bolometric fluxes \citep{2022ApJ...927...31T}, likely contributing to the poor quality sampling for these stars. We discuss some of these targets further in Section \ref{sec:mcmc_samp}.

\begin{deluxetable}{lccc}[htb!]
    \tablecolumns{4}
    \tablecaption{MCMC sampling offsets compared to SPORES catalog\label{ch4:tab:offsets}}
    \tablehead{\colhead{Parameter} & \colhead{Mean} & \colhead{Median} & \colhead{RMedS}}
    \startdata
    $\Delta T_{\rm eff}$ (K) & 17 ($0.4\%$) & 6 ($0.1\%$) & 6 ($0.1\%$) \\
    $\Delta R$ (R$_\odot$) & 0.01 ($1.4\%$) & 0 & 0 \\
    $\Delta$$[$Fe/H$]$ (dex) & 0.04 ($10.5\%$) & 0.01 ($2.3\%$) & 0.01 ($2.3\%$) \\
    $\Delta M$ (M$_\odot$) & 0.03 ($3.4\%$) & 0.03 ($3.1\%$) & 0.03 ($3.1\%$) \\
    $\Delta Age$ (Log yr) & 0.27 ($3.1\%$) & 0.12 ($1.2\%$) & 0.12 ($1.2\%$)
    \enddata
\end{deluxetable}

\subsection{Stellar Masses and Ages}\label{ch4:sec:ages_masses}

\begin{figure}[htb!]
    \centering
    \includegraphics[width=\textwidth]{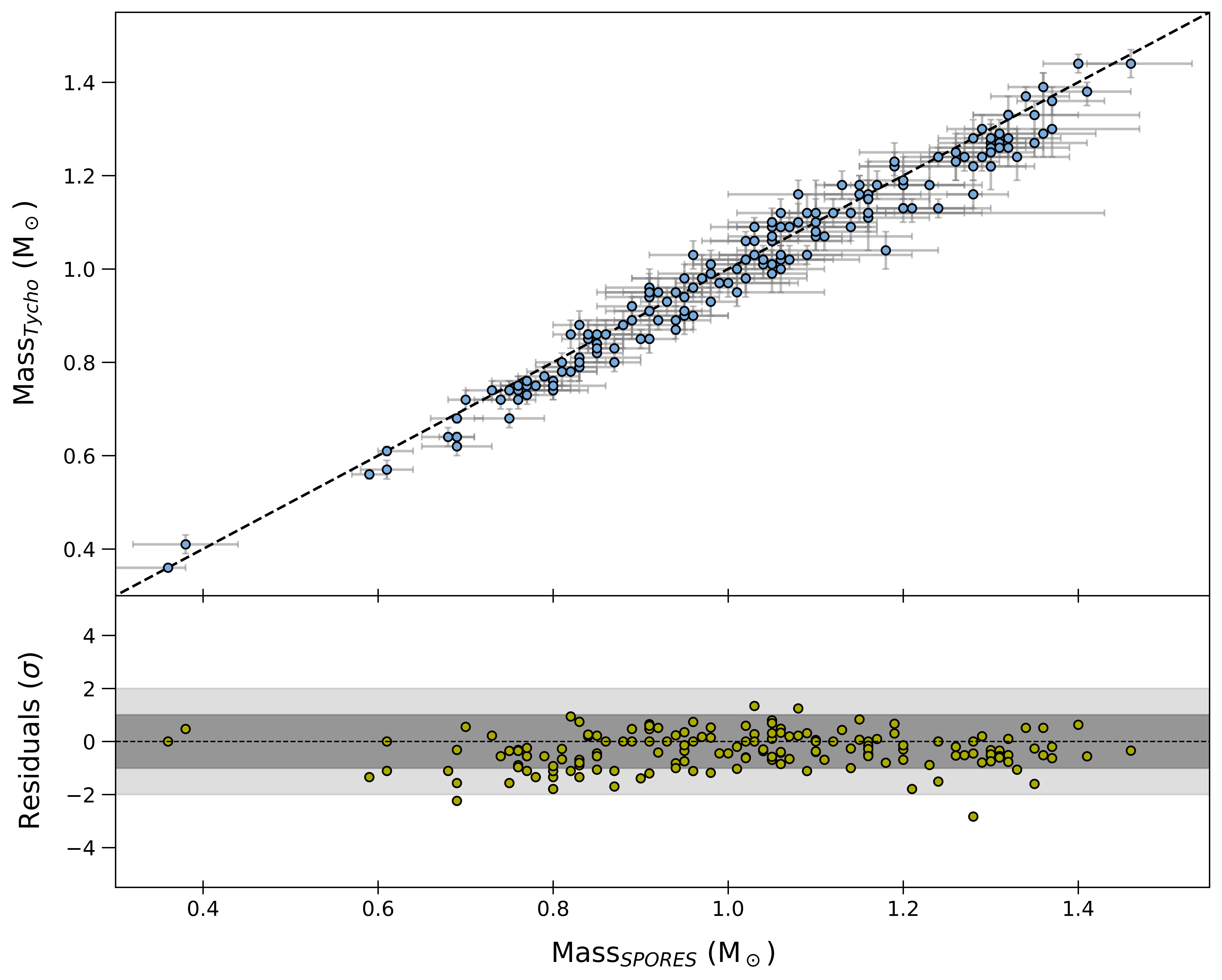}
    \caption{(top) Median masses from MCMC sampling of Tycho stellar model grid compared to masses from the SPORES catalog. The black dashed line represents a 1:1 correlation. (bottom) Residuals normalized by the quadrature sum of the uncertainties. The dark grey shaded region shows stars with $<1 \sigma$ discrepancy and the light grey $<2 \sigma$.}
    \label{ch4:fig:mass_comp}
\end{figure}

From our sampling of \teff, \rstar, and [Fe/H], we obtained posterior distributions for the masses and ages of the EMSL stars. We report the median and $68\%$ HPDI for the masses and ages of all 164 stars in Table \ref{ch4:tab:results}. Figure \ref{ch4:fig:mass_comp} compares the masses derived in this work to those in the SPORES catalog. The SPORES catalog similarly interpolated a stellar model grid to calculate masses and ages, but utilizing the MIST isochrone grid. Also, while the SPORES catalog used \teff\ and \rstar\ estimates from their SED fits as inputs, the authors used the [Fe/H] values from the EMSL instead of those obtained from the Hypatia catalog. We find a generally tight correlation, with all but two stars having a $<2 \sigma$ discrepancy as determined by the residuals normalized by the quadrature sum of the uncertainties. We compare the median uncertainties between both sets of masses in Table \ref{ch4:tab:median_unc}. We achieve a median uncertainty of 0.03 M$_\odot$ ($2.8\%$) from our sampling, compared to 0.05 M$_\odot$ ($4.6\%$) for the SPORES catalog. Overall, the two sets of masses agree well, with a RMedS offset of 0.03 M$_\odot$, well within the quadrature sum of the median uncertainties (0.06 M$_\odot$). The mean, median, and RMedS offsets for both the masses and ages are listed in Table \ref{ch4:tab:offsets}.

\begin{deluxetable}{ccccccccc}[htb!]
    \tablecolumns{9}
    \tablecaption{Derived masses, ages, and \chz\ metrics for EMSL stars\label{ch4:tab:results}}
    \tablehead{\colhead{HD} & \colhead{RA} & \colhead{DE} & \colhead{SpT} & \colhead{Tier} & \colhead{Age} & \colhead{Mass} & \colhead{\chz\ Metric} & \colhead{QF \tablenotemark{a}} \\
    \colhead{} & \colhead{deg} & \colhead{deg} & \colhead{} & \colhead{} & \colhead{Gyr} & \colhead{$M_\odot$} & \colhead{} & \colhead{}}
    \startdata
    HD 72673   & 128.2145653 & -31.5008514 & K1V    & C & $10.37^{+3.04}_{-1.25}$ & $0.74^{+0.02}_{-0.02}$ & 0.039 & 1 \\
    HD 72905   & 129.7987692 & 65.0209064  & G0.5V  & B & $2.25^{+0.79}_{-2.21}$  & $1.00^{+0.03}_{-0.03}$ & 0.163 & \\
    HD 74576   & 130.8251265 & -38.8823808 & K2.5V  & C & $8.72^{+4.88}_{-2.24}$  & $0.78^{+0.02}_{-0.03}$ & 0.045 & \\
    HD 75732 A & 133.1492129 & 28.3308208  & K0IV-V & C & $9.93^{+2.55}_{-2.49}$  & $0.90^{+0.02}_{-0.03}$ & 0.256 & \\
    HD 76151   & 133.5747796 & -5.4344595  & G2V    & C & $3.26^{+1.45}_{-1.13}$  & $1.01^{+0.02}_{-0.03}$ & 0.115 & \\
    HD 78366   & 137.2127933 & 33.8822184  & G0IV-V & C & $1.60^{+0.59}_{-1.57}$  & $1.07^{+0.03}_{-0.03}$ & 0.037 & \\
    HD 78154 A & 137.5981111 & 67.1340172  & F7V    & C & $2.80^{+0.40}_{-0.45}$  & $1.29^{+0.03}_{-0.04}$ & 0.865 & \\
    HD 82885 A & 143.9145925 & 35.8101325  & G8V    & B & $6.07^{+1.65}_{-1.58}$  & $0.97^{+0.02}_{-0.02}$ & 0.499 & \\
    HD 84117   & 145.5600675 & -23.9155672 & F9V    & A & $3.11^{+0.87}_{-0.69}$  & $1.12^{+0.04}_{-0.03}$ & 0.686 & \\
    HD 84737   & 147.1473806 & 46.0210074  & G0.5V  & B & $6.52^{+0.63}_{-0.63}$  & $1.12^{+0.03}_{-0.03}$ & 0.740 & 
    \enddata
    \tablecomments{Table \ref{ch4:tab:results} is published in its entirety in the machine-readable format. A portion is shown here for guidance regarding its form and content.}
    \tablenotetext{a}{ Quality flag on MCMC sampling.}
\end{deluxetable}

\begin{figure}[htb!]
    \centering
    \includegraphics[width=\textwidth]{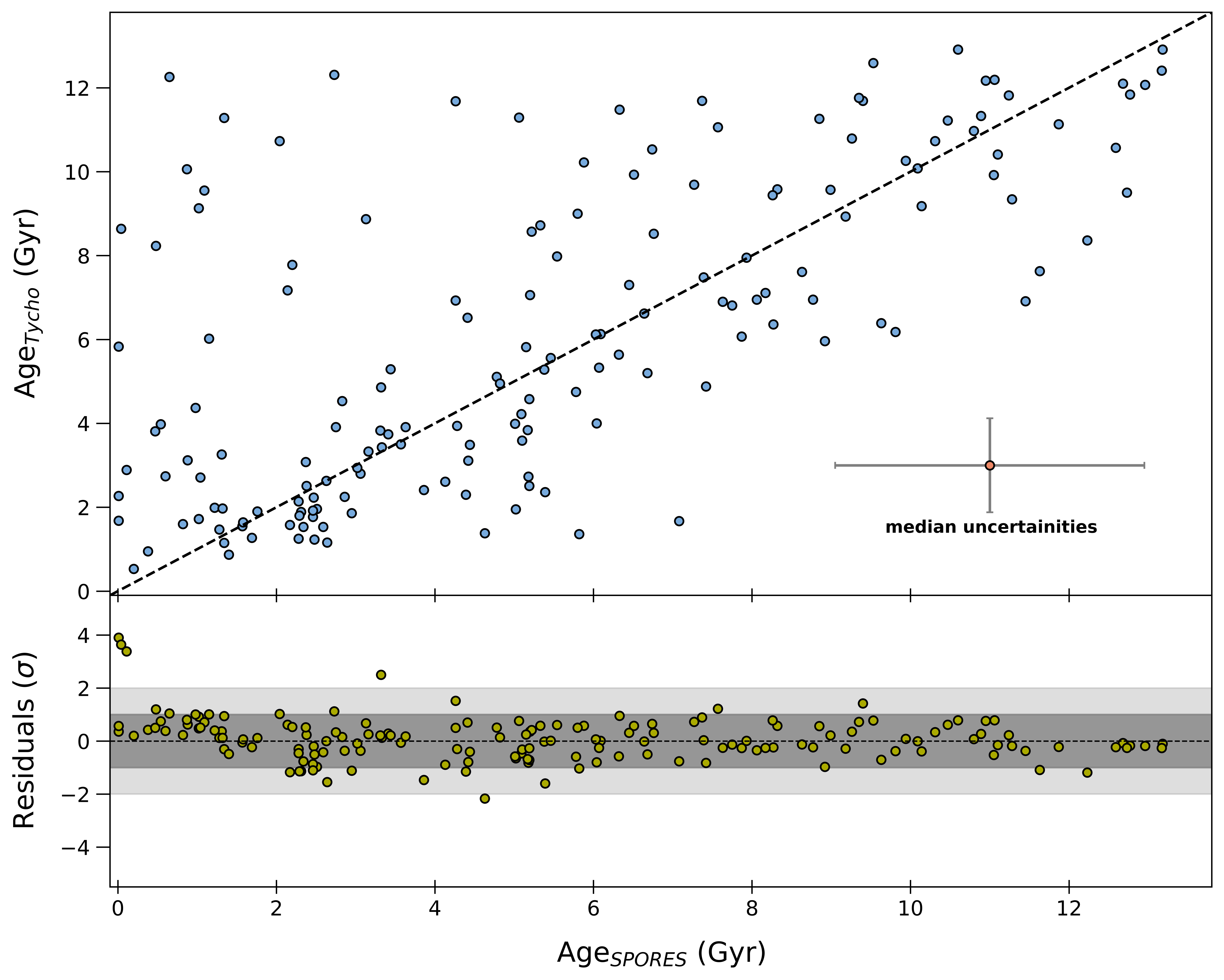}
    \caption{(top) Median ages from MCMC sampling of Tycho stellar model grid compared to ages from the SPORES catalog. The black dashed line represents a 1:1 correlation. Error bars are excluded for clarity. (bottom) Residuals normalized by the quadrature sum of the uncertainties. The dark grey shaded region shows stars with $<1 \sigma$ discrepancy and the light grey $<2 \sigma$.}
    \label{ch4:fig:age_comp}
\end{figure}

Figure \ref{ch4:fig:age_comp} compares the ages derived in this work to those in the SPORES catalog. Although there initially appear to be significant discrepancies between the two sets of ages, accounting for the uncertainties results in all but five stars having discrepancies $<2 \sigma$. This is indicative of the difficulty in determining ages from fits to stellar model grids, where uncertainties are generally $> 1$ Gyr and systematic differences between model grids are on the order of $\sim 20$\% \citep{2022ApJ...927...31T}. The Tycho ages systematically skew older for the youngest stars in the SPORES catalog. This is likely due to the EEP priors previously mentioned, where we restrict the EEP range to older than the ZAMS due to all of the EMSL stars being MS dwarfs and subgiants. We compare the median uncertainties between both sets of ages in Table \ref{ch4:tab:median_unc}. We achieve a median uncertainty of 0.09 dex ($0.9\%$) in log age from our sampling, compared to 0.19 dex ($2\%$) for the SPORES catalog. As with the masses, the overall ages of the two sets agree well, with a RMedS offset of 0.12 dex ($1.2\%$), which is within the quadrature sum of the median uncertainties (0.21 dex).

\begin{deluxetable}{lcc}[htb!]
    \tablecolumns{4}
    \tablecaption{Median mass and age uncertainties\label{ch4:tab:median_unc}}
    \tablehead{\colhead{Parameter} & \colhead{Median} & \colhead{Median$_{SPORES}$}}
    \startdata
    $\sigma_M$ (M$_\odot$) & 0.03 ($2.8\%$) & 0.05 ($4.6\%$) \\
    $\sigma_{Age}$ (Log yr) & 0.09 ($0.9\%$) & 0.19 ($2\%$)
    \enddata
\end{deluxetable}

\subsection{\chz\ Posterior Likelihoods}\label{ch4:sec:chz_prob}

From our sampling of the Tycho stellar model grid, we also obtained posterior distributions for the \chz\ boundaries. By comparing a range of orbital radii to the joint inner and outer \chz\ posterior samples, we calculate the \chz\ posterior likelihood distributions for all 164 stars in the EMSL. We convert the likelihood posteriors to a single \chz\ metric value using the method described in Section \ref{ch4:sec:metric}. The \chz\ metric results are reported in Table \ref{ch4:tab:results}.

\begin{figure}[htb!]
    \centering
    \includegraphics[width=\textwidth]{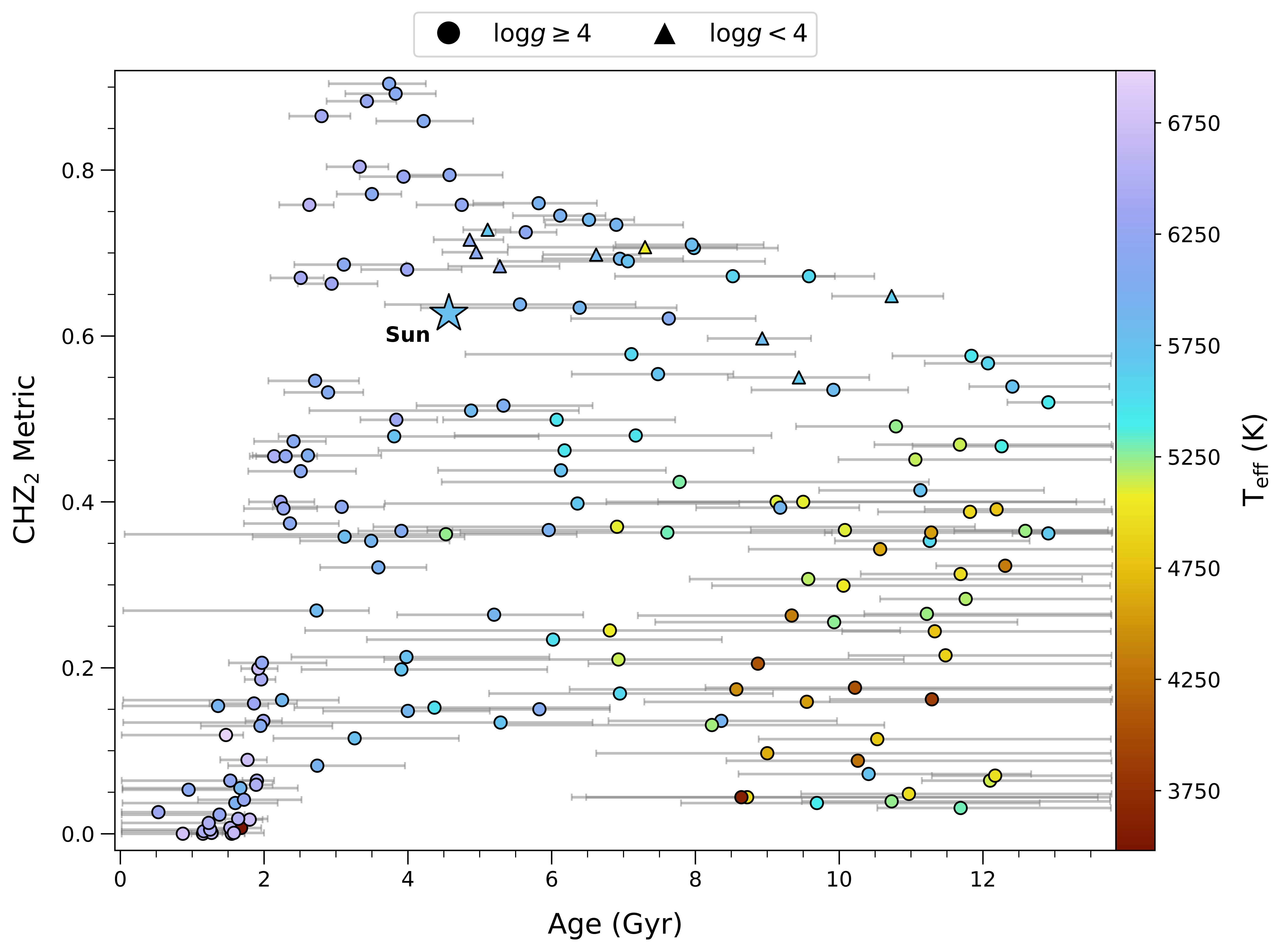}
    \caption{\chz\ metric vs. the age derived in this work for the 164 EMSL stars. Stars with log$g < 4$, determined by the Tycho derived mass and SPORES catalog \rstar, are marked by triangles to denote likely subgiants. Stars are colored according to the SPORES catalog \teff.}
    \label{ch4:fig:chz_metric}
\end{figure}

Figure \ref{ch4:fig:chz_metric} shows the \chz\ metrics with respect to the ages derived in this work and the SPORES catalog \teff. The \chz\ metric distribution begins at 0 for stars $<2$ Gyr in age, as these stars are not yet old enough to for the \chz\ to develop. The \chz\ metric then increases with age, peaking close to solar age with late-F to early-G stars typically having the highest values of $\sim 0.7-0.9$, before tapering off for the oldest stars. We mark stars with log$g < 4$, calculated via the mass derived in this work and the SPORES catalog \rstar, as these are likely to be subgiants. Although the sample of potential subgiants is small ($\lesssim 10$), these stars follow the same general trends as the sample of dwarf stars, albeit with lower \chz\ metrics than the peak as they tend to be older. We discuss these trends further in Section \ref{ch4:sec:disc_chz} along with how these metrics can influence prioritization and future work that is needed.

\subsection{Issues with Estimating Masses and Ages for the Lowest Mass Stars}\label{sec:mcmc_samp}

We determined masses, ages, and \chz\ metrics for the 164 stars in the EMSL through MCMC sampling of our Tycho stellar model grid, flagging the MCMC chains according to how well the inputs fit to the final sampling. We examine here several of the 10 stars with the poorest quality chains, where one or more input that fell outside the 99.7\% HPDI of the MCMC chains. These stars were primarily K- and M-type dwarfs, typically with sub-solar [Fe/H]. Stellar evolution models are known to struggle simulating low-mass stars, with \rstar\ underestimated by $\sim 5\%$ for stars $<0.7$ R$_\odot$ and \teff\ overestimated by $\sim 3\%$ for stars $<5000$ K \citep{2012ApJ...757..112B}. A major contributing factor is likely the exclusion of the effects of magnetic fields and starspots \citep[e.g.,][]{2008AA...478..507M,2020ApJ...891...29S}, which leads to an inflated \rstar\ and cooler \teff\ relative to predictions. This is exacerbated for lower metallicity stars as decreasing opacity causes the smaller model \rstar\ and higher model \teff.

While insufficient models likely explain the majority of the offset between the input and sampled \rstar\ and \teff, we also compare the SPORES catalog properties to those derived from interferometric angular diameters and bolometric luminosities. Six stars have an interferometric \rstar\ and \teff\ reported in the literature, with \teff\ generally consistent with those reported in the SPORES catalog. However, \rstar\ is found to be systematically overestimated in the SPORES catalog, with a median offset of 0.03 R$_\odot$ compared to interferometry for HD 88230 \citep{2012ApJ...757..112B}, HD 95735 \citep{2012ApJ...757..112B}, HD 201091 \citep{2012ApJ...757..112B}, HD 201092 \citep{2012ApJ...757..112B}, HD 209100 \citep{2009AA...505..205D}, and HD 217987 \citep{2012ApJ...757..112B}. As the posterior distributions skewed toward a smaller \rstar\ for these stars, the input \rstar\ may have contributed to the tension between the sampled \rstar\ and \teff.

\section{Discussion}\label{ch4:sec:disc}

Achieving the necessary signal-to-noise ratio (SNR = 5) in all wavelengths of interest to confidently characterize the atmospheres of terrestrial exoplanets in the HZs of Sun-like stars will require a steep observational cost. The EMSL, HabEx, and LUVOIR reports placed an upper exposure time limit of 60 days for suitable targets, implying that any given system could be observed for months for a complete survey of the HZ. If the ultimate goal of HWO is to detect biosignatures and constrain the prevalence of life in our galaxy, we must consider the potential for any HZ exoplanets to host detectable life. The \chz\ metric presented here provides an initial demonstration of how we can prioritize targets according to the orbital area around a star that is both accessible to an HWO-like mission and has remained in the HZ for a similar amount of time to Earth at the GOE. Systems with a higher \chz\ metric will have larger zones potentially capable of supporting temperate terrestrial exoplanets and may be more desirable for observations.

\subsection{\chz\ Metric}\label{ch4:sec:disc_chz}

\begin{figure}[htb!]
    \centering
    \includegraphics[width=\textwidth]{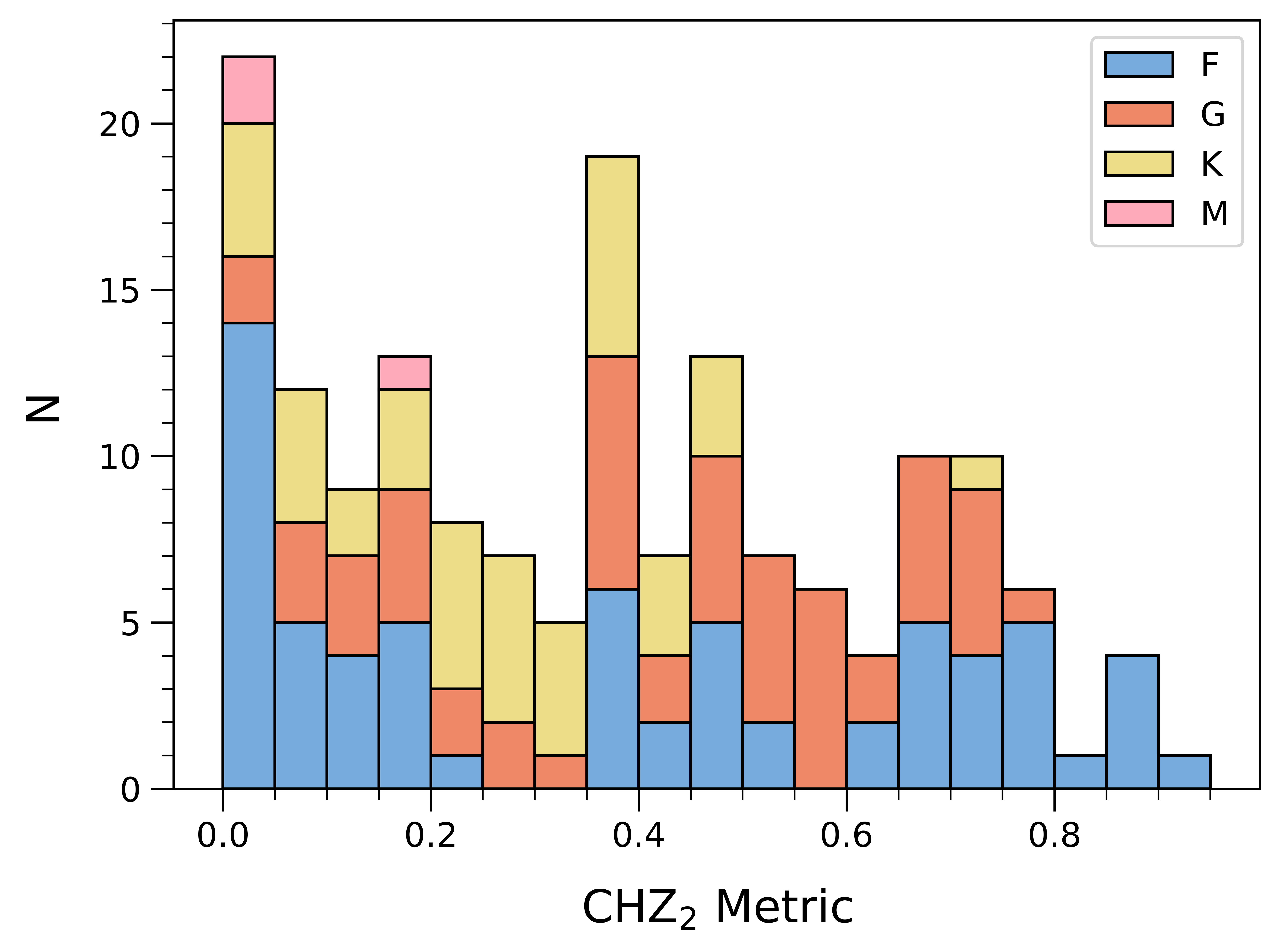}
    \caption{Distribution of \chz\ metric values separated by spectral type for the 164 EMSL stars.}
    \label{ch4:fig:chz_metric_hist}
\end{figure}

The \chz\ metrics show expected trends with respect to age and spectral type. Older stars in this sample tend to have slightly lower \chz\ metrics, likely due to two factors. The \chz\ begins to contract as stars reach the later MS, with more rapid changes in luminosity as hydrogen is depleted in the core. Further, due to the combination of a slower rate of evolution and larger relative uncertainties on stellar radius, K dwarfs are biased towards older derived ages. This biases the \chz\ metric toward lower values at older ages. For F, G, and K stars in the EMSL, the median values we derive for the \chz\ metric are 0.393, 0.467, and 0.264, respectively. Although K dwarfs have longer main sequence lifetimes, the size of the HZ scales with $(\frac{L}{L_\odot})^{0.5}$ and hence results in a smaller \chz. The opposite is true for early-F dwarfs, with late-F and G dwarfs striking a balance between these two with both long MS lifetimes and large HZs. This is portrayed in Figure \ref{ch4:fig:chz_metric_hist}, with F and G dwarfs dominating the highest \chz\ metric values.

Although early-F dwarfs will reach a higher peak \chz\ width, this is short-lived ($\sim 100$ Myr) due to shorter stellar lifetimes and rapid evolution. Stars earlier than $\sim {\rm F3-F4}$ show a steep drop in \chz\ metric, with a median value of 0.053. Similarly, a drop is seen for stars with $T_{\rm eff} \gtrsim 6600$ K or ${\rm mass} \gtrsim 1.3$ $M_\odot$. These results are for a relatively small sample of stars ($\sim 10$), but are an indicator of the cut-off point for long-lived CHZs. Although these stars may gain a \chz\ at some point during their evolution, we tentatively state that stars earlier than $\sim {\rm F3}$ or hotter than $\sim 6600$ K are unlikely to have a \chz\ at the time of observation. A definitive cut-off point, where the \chz\ ceases to exist at any point during the stellar lifetime, can be determined through evolution models. For models with ${\rm [M/H]} \leq 0$ in the Tycho grid, a \chz\ is never obtained at masses above 1.57 $M_\odot$ during the MS, but a \chz\ may still exist during the subgiant branch or for super-solar metallicities. Since the Tycho grid only extends up to 1.6 $M_\odot$ to encompass the EMSL stars, all masses still have some models with a \chz. An extended grid will likely show a true cut-off point between $1.6-1.7$ $M_\odot$ as the stellar lifetime drops below 2 Gyr.

Notably, the \chz\ metric peaks between $3-4$ Gyr for late-F dwarfs (Figure \ref{ch4:fig:chz_metric}). Therefore, stars slightly younger and more massive than the Sun may be the most likely to have the widest zones of continuous habitability. This coincidentally synergizes with the results from \citet{2023IJAsB..22..272M}, in which the authors combine a minimal mantle and atmospheric model to study the evolution of the CHZ for various planetary parameters and stellar types. In their model, the IHZ expands as the luminosity increases and the OHZ contracts as geologic activity, and hence CO$_2$ outgassing, decreases with time. This results in planets becoming geologically dead within $\sim 5-7$ Gyr for those between $0.5-1$ M$_\oplus$, after which the planets rely solely on their host star's luminosity to maintain habitable surface temperatures. The authors speculate that the average age for habitable planets is therefore younger than the age of the Earth, with only super-Earth's maintaining habitable conditions for longer periods.

One may expect subgiants in this sample to have significantly lower \chz\ metrics than dwarfs as they leave the main sequence and begin to rapidly evolve. However, the luminosity remains relatively constant on the subgiant branch, only changing once the star nears the RGB. \teff\ decreases significantly throughout the subgiant branch, reaching $\sim 4000$ K for a solar twin at the base of the RGB. This will have a slight effect on the HZ boundaries by pushing the HZ further out as the Wien peak shifts towards the infrared, but the overall shift is marginal relative to the luminosity of the star at the TAMS. The impact this shift in peak wavelength would have on life is less clear, but increased interest in the potential for life on exoplanets has spurred further research. A recent study \citep{2024IJAsB..23..e18V} explored the effects of two photosynthetic organisms grown under a solar spectrum and a simulated K dwarf spectrum (4,300 K). Garden cress \textit{Lepidium sativum}, a common garden plant, exhibited comparable growth and photosynthetic efficiency, a measure of the effectiveness with which an organism converts photons to chemical energy, under both spectra. Surprisingly, the cyanobacterium \textit{Chroococcidiopsis} exhibited increased culture growth and photosynthetic efficiency under the K dwarf spectrum. Although further research is needed to apply a broad consensus for life on Earth, these results are encouraging for the prospect of life on other worlds surviving the spectral shift during the subgiant branch.

Overall, the five highest ranked stars according to the \chz\ metric are HD 102870 (F9V), HD 9826 A (F8V), HD 126660 A (F7V), HD 78154 A (F7V), and HD 5015 (F8V), all with estimated ages slightly younger than the Sun ($\sim 3-4$ Gyr). However, all of these stars are in EMSL Tier C. HD 102870, HD 126660 A, and HD 78154 A have known warm dust disks or binary companions at $<5$" separations, which will negatively impact the sensitivity of direct imaging observations. Taking into account the presence of known disks and close binaries, the five highest ranked stars are HD 9826 A (F8V, Tier C), HD 5015 (F8V, Tier C), HD 142860 (F6V, Tier A), HD 187691 A (F8V, Tier B),  and HD 19373 (G0V, Tier A). Similarly, these stars are close in age to the Sun ($\sim 3-6$ Gyr) and later spectral types with three still in lower EMSL tiers. When calculating the \chz\ metric, we assume that any exoplanet outside the IWA will be observable by HWO. However, these exoplanets may still be too dim to observe. The EMSL used 12 cases to determine the observability of HZ exoplanets for a given system, calculating the exoplanet brightness and planet-to-star contrast ratio for a multiple planet radii (1 and 1.4 $R_\oplus$), orbital phases (90$^\circ$ and 63.3$^\circ$), and HZ locations (1, 1.31, and 1.55 AU) scaled by the host star luminosity. Since the HZ scales as $(L/L_\odot)^{0.5}$, exoplanets in the HZs of F-type stars will reflect a smaller portion of their host stars light than for later-type stars. This leads to numerous otherwise suitable F-type stars being relegated to lower EMSL tiers, with the hypothetical exoplanet cases not reaching the exoplanet brightness and planet-to-star contrast ratio limits for Tier A.

While it may be useful to include assumptions about exoplanet brightness in future iterations of the \chz\ metric, the method for calculating HZ distance in the EMSL is likely injecting a bias against hotter stars. For standard F9 (6050 K, 1.66 $L_\odot$) and F1 (7020 K, 6.17 $L_\odot$) dwarfs, representative of the range included in the EMSL, the \citet{2014ApJ...787L..29K} IHZ distances are 1.2 and 2.2 AU, respectively. If we instead scale the 1 AU IHZ distance for Earth by $(L/L_\odot)^{0.5}$, this results in IHZ distances of 1.29 and 2.48 AU, corresponding to reductions in exoplanet brightness of 14\% and 28\%. This example is also a significant underestimate of the total difference to expect, since the change in HZ distance is due to the relationship between the stellar spectrum and planetary albedo. The EMSL assumed a geometric albedo of 0.2 for all 12 exoplanet test cases, but Rayleigh scattering becomes increasingly important as \teff\ increases. Higher flux at shorter wavelengths increases the planetary albedo and pushes the HZ closer in. Similarly, the brightness of exoplanets orbiting cooler hosts will be overestimated, with lower albedos and more distant HZs when factoring in \teff. Due to these different assumptions about the exoplanet brightness and HZ boundaries, the \chz\ metric tends to rank earlier spectral types higher than in the EMSL.

Ultimately, the \chz\ metric would be best utilized as a term in an EMSL-like framework. When comparing two targets, a small difference in the \chz\ metric would not justify prioritizing one over the other if the required exposure time would increase significantly. Similarly, the \chz\ metric would provide an easy method for distinguishing between targets with similar observational costs if one system's likelihood of continuous habitability provides increased scientific justification. A more integrated approach could involve replacing the classical HZ with the \chz\ when investigating test cases for exoplanet brightness, taking into account the \chz\ likelihood instead of a binary determination of if the exoplanet would be observable.

\subsection{Future Work}\label{ch4:sec:future}

In this work we have only focused on using the Tycho grid of stellar evolution models, combined with the HZ prescriptions of \citet{2014ApJ...787L..29K} and \citet{2023AA...679A.126T}, to estimate ages, masses, and \chz\ posterior likelihood distributions for stars in the EMSL. There are a plethora of stellar model grids that exist, with some taking into account effects of rotation \citep[e.g., YREC,][]{2020ApJ...888...43C} and magnetic fields \citep[e.g., SPOTS,][]{2020ApJ...891...29S}. Considering our difficulties with sampling for the lowest mass K and M dwarfs and the known systematic offsets between model grids \citep[e.g.,][]{2022ApJ...927...31T}, it will be essential to compare to predictions from a variety of stellar models. While the \citet{2014ApJ...787L..29K} HZ prescriptions are some of the most commonly used in the exoplanet community, a large number of HZ prescriptions have been reported in the literature over the past two decades \citep[e.g., see][and references therein]{2017ApJ...837..107W}. Considering the limitations of 1D climate models in realistically simulating planetary atmospheres and the various planetary parameters that can impact the climate, comparisons with HZ prescriptions determined with 3D GCMs and alternative planetary configurations will better inform target prioritization for HWO. A benefit of using \texttt{kiauhoku} is that we can easily compare offsets between stellar model grids for derived masses and ages. With the addition of our module for computing \chz\ posterior distributions, we will also be able to compare offsets between grids for the \chz\ metrics.

Stellar ages are notoriously difficult to constrain for field dwarfs, especially low-mass K and M dwarfs with long main sequence lifetimes. Alternative methods to fitting surface properties to stellar model grids often provide more precise estimates, such as through asteroseismology \citep[e.g.,][]{2022AJ....163...79H} or empirical relations between age and rotation \citep[e.g.,][]{2010ApJ...721..675B}, activity \citep[e.g.,][]{2018AA...619A..73L}, or elemental abundances \citep[e.g.,][]{2015AA...579A..52N}. We plan to incorporate age estimates from these sources in the future to enable better constraints on the current location of the \chz.

\section{Conclusion}\label{ch4:sec:conc}

Searching for terrestrial HZ exoplanets around Sun-like stars and probing their atmospheres for biosignatures through direct imaging will be a daunting task, but represents our best chance at constraining the prevalence of life in our galaxy. Considering the potential for life to have made a detectable impact on the atmosphere presents a means to prioritize targets in the lead-up to future missions. In this work, we combined stellar evolution models and published HZ prescriptions to calculate a \chz\ metric for the 164 stars in the EMSL, which may be the most suitable targets for the future HWO direct imaging mission. The \chz\ metric is a proxy for the width of the \chz, but is weighted by the likelihood that each orbital radius is within the \chz. The metric demonstrates a method for ranking targets according to both the likelihood that exoplanets will have remained in the HZ for a similar time-frame to Earth and the accessibility of the \chz\ for an HWO-like mission. Here we provide a summary of our results:

\begin{itemize}
    \item We report \chz\ metrics for the 164 stars in the EMSL. Our results show that the \chz\ metric peaks between $3-4$ Gyr for late-F and early-G dwarfs, indicating that these may be excellent targets for habitable exoplanet searches. However, a comparison to the tiered rankings in the EMSL showed we tend to give higher prioritization to F dwarfs, which have the largest CHZ$_2$s.
    \item We tentatively find that stars earlier than $\sim {\rm F3}$ or hotter than $\sim 6600$ K are unlikely to have a \chz\ at the time of observation, but a larger sample is needed to confirm this result.
    \item We also derived isochronal masses and ages for the EMSL stars. Comparison of our masses and ages to those in the SPORES catalog \citep{2024ApJS..272...30H}, from which we adopted their uniformly derived \teff\ and \rstar\ for our calculations, showed generally good agreement. The RMedS offsets are comparable to the median uncertainties, but with improved precision for our results.
    \item We developed an open-source module for the stellar model grid interpolation code \texttt{kiauhoku} \citep{2020ApJ...888...43C}, which combine to enable simultaneous sampling of stellar age, mass, and \chz\ boundaries for any stellar model grid.
    \item Future work is needed to compare \chz\ metric results between stellar model grids and for various HZ prescriptions. Additionally, we struggled to reliably model a number of K and M dwarfs, a well known problem for stellar evolution models \citep[e.g.,][]{2012ApJ...757..112B}. The addition of the effects of magnetic fields and starspots to Tycho or the use of a model grid that already accounts for some of these effects \citep{2020ApJ...891...29S} may be necessary to achieve reliable sampling for these stars.
\end{itemize}

\begin{acknowledgements}
    The authors would like to thank the anonymous reviewer and Eric Mamajek for their insightful comments, which greatly improved the content of this manuscript. The results reported herein benefited from collaborations and/or information exchange within NASA’s Nexus for Exoplanet System Science (NExSS) research coordination network sponsored by NASA’s Science Mission Directorate (Grant 80NSSC23K1356, PI Steve Desch).
\end{acknowledgements}

\appendix
\restartappendixnumbering

\section{Alternative Red Giant Branch Cutoff}\label{sec:eep}

\begin{figure}[htb!]
    \centering
    \includegraphics[width=\textwidth]{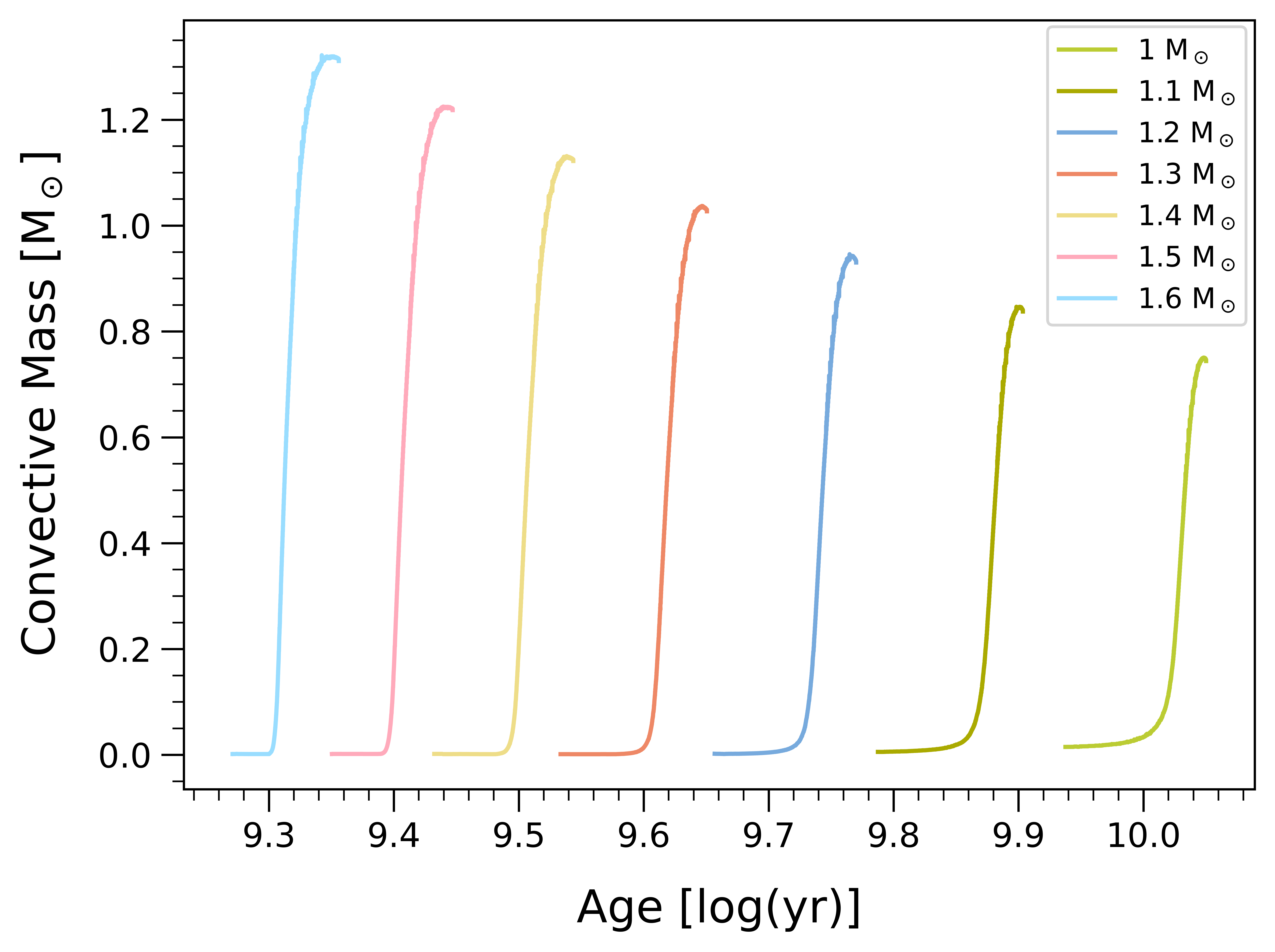}
    \caption{Evolution of the total convective mass for $1-1.6$ M$_\odot$, solar metallicity Tycho stellar evolution tracks. Here, tracks start from the base of the RGB and end just after the maximum convective depth is reached during the first dredge-up. The maximum convective mass provides a common final EEP for all tracks that reach the RGB in our model grid.}
    \label{ch4:fig:conv_mass}
\end{figure}

In Section \ref{ch4:sec:kiauhoku}, we use stellar model grid interpolation to determine the stellar age, mass, and \chz\ posterior likelihood distributions for all stars in the EMSL. When interpolating between tracks, stellar age is not an optimal dimension for comparison. Depending on the initial mass and metallicity, consecutive tracks can vary substantially in the lifetime of each evolutionary phase and which phases are reached. In order to simplify and improve the accuracy of interpolation, the original tracks output by the stellar evolution model are transformed onto a uniform basis, with an equal number of steps for each evolutionary phase. We transform the Tycho stellar evolution catalog tracks to an equivalent evolutionary phase (EEP) basis, using the EEP definitions of \citet{2016ApJS..222....8D}. The main difference is that we do not go up to or past the tip of the red giant branch (RGB). We only need to simulate up to the end of the subgiant branch for the EMSL stars, so we stop the tracks at the maximum extent of the first convective dredge-up. The first dredge-up is an event in the first half of the RGB where the convective envelope extends deep enough to bring the products of hydrogen burning to the surface. The maximum convective depth provides an easily distinguishable feature when examining the total convective mass in the stellar interior (Figure \ref{ch4:fig:conv_mass}) and we apply this as the final EEP for all tracks in our grid that reach the RGB. We recommend this EEP for studies focused on earlier phases of evolution to avoid the large computational time required to reach the helium flash at the tip of the RGB.

\section{Spectral Types in the EMSL}\label{sec:spt}

Due to an error in the tabulation of spectral types for the EMSL paper, there are mismatched spectral types reported in the EMSL and SPORES catalog for a significant number of G-type stars. We have performed a literature search and retrieved spectral types for all 164 stars, adopting new spectral types for 18 stars. Full spectral types, including additional specifiers, for those 18 stars are shown in Table \ref{ch4:tab:spt}, along with references and the spectral types reported in the EMSL. Spectral types reported in Table \ref{ch4:tab:results} include these updates, but are abbreviated to remove additional specifiers. A future version of the EMSL will include the updated spectral types for these stars (Eric Mamajek, \textit{priv. comm.}).

\begin{deluxetable}{cccc}[htb!]
    \tablecolumns{4}
    \tablecaption{Updated spectral types for EMSL stars\label{ch4:tab:spt}}
    \tablehead{\colhead{HD} & \colhead{SpT$_{\rm EMSL}$} & \colhead{SpT$_{\rm lit}$} & \colhead{Ref. \tablenotemark{a}}}
    \startdata
    HD 82885 A  & G9-IV-V Hdel1  & G8+V              & 1 \\
    HD 84737    & G0(V)          & G0.5Va            & 2 \\
    HD 86728 A  & G4IV           & G3Va Hdel1        & 2 \\
    HD 95128    & G1.5IV-V Fe-1  & G1V               & 3 \\
    HD 131156 A & G8-V           & G7V               & 1 \\
    HD 136352   & G2.5V Hdel1    & G2-V              & 4 \\
    HD 142373   & G0V Fe+0.4     & G0V Fe-0.8 CH-0.5 & 1 \\
    HD 143761   & G0IV           & G0V               & 1 \\
    HD 146233   & G3+V           & G2V               & 1 \\
    HD 160691   & G3V            & G3IV-V            & 4 \\
    HD 182572   & G7IV-V         & G7IV Hdel1        & 2 \\
    HD 190360   & G7V            & G7IV-V            & 4 \\
    HD 189567   & G2V Fe-1.0     & G2V               & 4 \\
    HD 190248   & G8IV-V         & G8IV              & 4 \\
    HD 207129   & G0VmF2         & G0V Fe+0.4        & 4 \\
    HD 20766    & G2IV           & G2.5V Hdel1       & 2 \\
    HD 30495    & G1.5V(n)       & G1.5V CH-0.5      & 4 \\
    HD 34411    & G1.5V          & G1V               & 4
    \enddata
    \tablenotetext{a}{ References: 1. \citet{2003AJ....126.2048G} 2. \citet{1989ApJS...71..245K} 3. \citet{2001AJ....121.2148G} 4. \citet{2006AJ....132..161G}}
\end{deluxetable}


\bibliography{main.bib}{}
\bibliographystyle{aasjournal}



\end{document}